\documentclass[preprint,12pt]{elsarticle}
\let\today\relax
\makeatletter
\def\ps@pprintTitle{%
    \let\@oddhead\@empty
    \let\@evenhead\@empty
    \def\@oddfoot{%
        \footnotesize\itshape
        \parbox{\textwidth}{%
            This is an Accepted Manuscript version of the following article, accepted for publication in \@journal. Mármol-Romero, A. M., García-Vega, M., García-Cumbreras, M. Á.,  Montejo-Ráez, A. (2024). An Empathic GPT-Based Chatbot to Talk About Mental Disorders With Spanish Teenagers. International Journal of Human–Computer Interaction, 41(7), 3957–3973. https://doi.org/10.1080/10447318.2024.2344355. It is deposited under the terms of the Creative Commons Attribution-NonCommercial-NoDerivatives License, which permits non-commercial re-use, distribution, and reproduction in any medium, provided the original work is properly cited, and is not altered, transformed, or built upon in any way.
            \hfill\today
        }%
    }%
    \let\@evenfoot\@oddfoot
}
\makeatother

\usepackage{amssymb}

\usepackage{float}
\usepackage{inconsolata}
\usepackage{graphicx}
\usepackage{caption}
\usepackage{subcaption}
 \usepackage{booktabs}
\usepackage{multirow}
\usepackage{tabularx}
\usepackage{pgf-pie}
\usepackage{xcolor}
\usepackage{url}
\journal{International Journal of Human-Computer Interaction}

\begin{document}

\begin{frontmatter}

\title{An empathic GPT-based chatbot to talk 
about mental disorders with Spanish teenagers}

\author[inst1]{Alba María Mármol-Romero}
\author[inst1]{Manuel García-Vega}
\author[inst1]{Miguel Ángel García-Cumbreras}
\author[inst1]{Arturo Montejo-Ráez}

\affiliation[inst1]{organization={CEATIC, Universidad de Jaén},
            addressline={Las Lagunillas s/n}, 
            city={Jaén},
            postcode={23071}, 
            country={Spain}}

\begin{abstract}
This paper presents a chatbot-based system to engage young Spanish people in the awareness of certain mental disorders through a self-disclosure technique. The study was carried out in a population of teenagers aged between 12 and 18 years. The dialogue engine mixes closed and open conversations, so certain controlled messages are sent to focus the chat on a specific disorder, which will change over time. Once a set of trial questions is answered, the system can initiate the conversation on the disorder under the focus according to the user's sensibility to that disorder, in an attempt to establish a more empathetic communication. Then, an open conversation based on the GPT-3 language model is initiated, allowing the user to express themselves with more freedom. The results show that these systems are of interest to young people and could help them become aware of certain mental disorders.
\end{abstract}


\end{frontmatter}

\section{Introduction}
Specific population groups such as children and adolescents have been identified by the World Health Organization (WHO) as being at particular psychological risk. According to the WHO's report \cite{world2022world}, about one in eight people in the world live with a mental disorder. Distress, harassment, abuse, manipulation, extortion, bullying, incitement to suicide or misbehaviours\footnote{https://www.who.int/news-room/fact-sheets/detail/adolescent-mental-health} are attacks that have been leveraged by the new forms of social interaction promoted by the Internet.

Depression and anxiety are increasing faster than ever, and suicide rates are rising, as revealed by recent data. Suicide is likely to become a more pressing concern as the COVID-19 pandemic spreads and has longer-term effects on the general population, the economy and vulnerable groups \cite{life11111188, hawes2021increases}. Preventing suicide, therefore, needs urgent consideration by addressing emotional problems. Compared to estimates before the epidemic, reports already indicate a 3-fold increase in the prevalence of experience of mood and anxiety swings \footnote{https://www.who.int/news-room/fact-sheets/detail/suicide}. The intensive presence of youth in social networks could make us think of solutions to monitor their behaviour, as human language is the main transmission medium involved in social interaction. Human language technologies can help us to build more confident environments, by using systems that continuously analyze the messages of a user in these platforms to predict the risk of suffering, for instance, depression. The use of social networks is fully correlated with the prevalence of these problems \cite{cdc2023} but the willingness of young people to allow an external system that could monitor their private conversations carries many problems in terms of privacy and legal aspects.

Our proposal explores an intermediate solution: to make a chatbot available on a well-known platform (Telegram) and let young people chat with it about different disorders. In this way, we can help them to be more aware of the existence of these disorders, share strategies to deal with them and compile a corpus of messages for further research on how valid is human language as an instrument for the early detection of certain mental disorders.

The paper is organized as follows: first, the state of the art in related technologies and experiences is provided concerning dialogue systems in general and chatbots for mental health care in particular. In Section 3, the methodology followed in our study is explained, introducing the recruitment process, and the design and construction of the bot with details of their main components and purpose. In Section 4 the results of its use are depicted quantitatively (usage statistics) and qualitatively (analysis of the answers to the survey by participants). In Section 5, we discuss the findings. Finally, in the Conclusions section, the main contributions of our study are highlighted with some proposals for future work. 

\section{Related work}
\subsection{Dialogue systems}
Dialogue systems, also known as conversational agents, are computer programs designed to engage in natural language conversations with human users. These systems can be categorized into several types, such as task-oriented dialogue systems and open-domain dialogue systems \cite{3166058}. This is an area of research that dates back to 1966, with the creation of the ELIZA system \cite{weizenbaum1966eliza}. It simulated conversation by using pattern matching and substitution methodology that gave users an illusion of understanding on the part of the program. Parry was a 1972 computer program designed to simulate a person with paranoid schizophrenia \cite{colby1981modeling}. It was designed to demonstrate that computer programs could pass the Turing test by being indistinguishable from a human. 

Many systems, architectures and approaches have been studied and proposed since then. Recent advances in Deep Learning (DL) and Natural Language Processing (NLP) have led to significant progress in dialogue systems. The introduction of the Transformer architecture \cite{vaswani2017attention}, and encoder-decoder neural network that has emerged as a prominent solution paradigm in the NLP. One of the most notable developments in recent years has been the emergence of large pre-trained language models based on the transformer model. Some of them have focused on the encoder part like BERT \cite{devlin-etal-2019-bert} and its derivatives. But the Large Language Models (LLM) offering pre-trained instances on the decoder part, the auto-regressive one, such as GPT-3 \cite{brown2020language}, have shown impressive performance in generating coherent and contextually relevant responses. These models have improved in the area of dialogue systems research to change them forever, due to their superior capabilities in language understanding. In fact, conversational-oriented models are becoming the most advanced ones, being some of the most revolutionary ones the well-known ChatGPT (technically known as GPT3.5) \cite{zhou2023comprehensive}, LaMDA \cite{51115} by Google, LLaMA \cite{touvron2023llama} by Meta,  and the brand new GPT-4 \cite{gpt4} by OpenAI. 

Another area of research that has shown promise is the use of reinforcement learning \cite{NEURIPS2021_7f489f64} to train dialogue agents. Reinforcement learning allows agents to learn from their interactions with users and optimize their behaviour accordingly. This approach has been applied in OpenAI models, like GPT-3, ChatGPT, and GPT-4, as a strategy to fine-tune the models in aspects like avoiding biased or stereotyped answers. Instead, Google's conversational models, like LaMDA, focus more on enriching knowledge by using some sort of ``extensions'' like information retrieval engines or calculators. This is an approach that is gaining attention from other companies, like Meta AI Research with its recent Toolformer \cite{schick2023toolformer}. 

Finally, research in multimodal dialogue systems (like GPT-4 or Microsoft's KOSMOS-I \cite{huang2023language}), which can integrate multiple sources of information such as text, speech and visual cues, has shown great potential for improving the naturalness and effectiveness of dialogue systems.

\subsection{Chatbots for mental health care}
In medicine in general, chatbots are becoming increasingly popular, for example, chatbots have recently been developed to solve health-related questions \cite{siddique2021machine,chow2023developing}, which could be extended to mental health issues. But nowadays, chatbots in particular are focusing on mental health issues \cite{vaidyam2019chatbots}.

Chatbots for mental disorders have shown great potential as a tool for providing mental health support to individuals in need \cite{balaskas2021ecological}. These chatbots use NLP and Machine Learning (ML) techniques to provide personalized support and guidance to users \cite{ahmed2023chatbot}. 

But what is more remarkable is that the latest research findings reveal that people often perceive artificial intelligence such as Alexa, Siri, or Google Assistant as companions or even members of their extended family \cite{skjuve2021my}. This highlights a notable trend in human-AI relationships where these digital entities transcend mere tools and take on relational roles in users' lives. In light of this, the study incorporates the use of the ``self-disclosure" technique, a proven strategy for fostering trust and encouraging individuals to articulate their concerns with a chatbot \cite{luo2020self} \cite{lee2020designing}. Several studies emphasise that the use of this technique is an important factor in establishing good human-machine communication \cite{skjuve2022longitudinal, skjuve2021my}. This strategy refers to the act of revealing personal information or details about oneself to others. It's a communication technique where individuals share thoughts, feelings, experiences, or opinions about themselves intending to foster understanding, trust, and connection with others \cite{derlega1993self, altman1973social}. Self-disclosure manifests in various contexts, encompassing personal relationships, therapy sessions, social interactions, and even online communication, and prior research has identified it as a potential path to mental wellness. This concept aligns with interpersonal communication theories, exemplified by the Social Penetration Theory (SPT) \cite{carpenter2015social}, which asserts that self-disclosure is pivotal in relationship development. As individuals progressively disclose more about themselves, the relationship's depth increases, fostering heightened intimacy and understanding. In addition, SPT provides three layers with different depths of self-disclosure. In this work, we apply the first layer (peripheral layer) where we talk about general things like age, name, etc. and the third layer (central layer) where we talk about feelings and deeper issues. However, until now, no prototype chatbot has been developed that uses self-disclosure to engage in conversations on sensitive topics for Spanish. So we present the chatbot as a teenager and it is vulnerable, trusting and revealing.

In addition, extensive studies have determined that gender and self-disclosure are two related variables \cite{hill1987gender, yu2019almost}, so we make it easy for users to choose the gender of the bot they are most comfortable with.

The psychological foundation in the use of chatbots is Cognitive-Behavioral Therapy (CBT) \cite{hollon2013cognitive}, which is a type of psychotherapy that focuses on helping individuals understand and modify their thoughts, behaviours and emotions. CBT works on the basis that our thoughts, emotions, and behaviours are all connected and can influence one another. Through CBT, individuals learn how to identify and challenge distorted or unhelpful thoughts, develop more effective coping strategies for managing stress and difficult emotions, and learn how to make more positive changes in their behaviours.

One notable example of a chatbot for mental disorders is Woebot\footnote{https://woebothealth.com/try-woebot/}, which uses cognitive-behavioural therapy techniques to provide support for anxiety and depression. Woebot uses NLP to understand users' emotions and provide tailored feedback and guidance in brief, daily conversations and mood tracking. It has been found useful for CBT-related therapies \cite{7785}. 

Another example is Wysa\footnote{https://www.wysa.com/} \cite{inkster2018empathy}, a chatbot that uses a combination of CBT, mindfulness, and Dialectical Behaviour Therapy (DBT) to provide support for a range of mental health issues. Wysa also uses NLP to understand users' emotions and provide personalized support. 

Research in this area has also explored the use of chatbots for mental health screening and assessment \cite{ahmed2023chatbot}, as well as for providing psychoeducation and self-help resources \cite{abd2019overview}. However, experts suggest they are not a substitute for professional mental health care and should be used in conjunction with other forms of treatment  \cite{vaidyam2019chatbots}.


\section{Methodology}
\label{sec:methodology}
The BigHug project\footnote{https://bighug.ujaen.es/} represents a novel approach for the early detection of mental disorders and misbehaviour (depression, anxiety, eating disorders, gambling addiction, suicidal ideation, and cyberbullying) in online social networks. This project was developed in different steps: (1) recruitment of subjects, interviews and clinical evaluations, (2) chatbot intervention and (3) analysis and system evaluation. This project was conducted by the University of Jaén (Spain) and involved the joint effort of researchers from the Computer Science and Psychology domains.

\subsection{Recruitment}
The original goal of the project was to build a tool for the early detection of mental disorders and misbehaviour by monitoring social networks. To this end, a recruitment process was initiated with the collaboration of several secondary schools in the province of Jaén, in the South of Spain.
Young students from ten different educational centres and ages between twelve and eighteen years old participated voluntarily. Psychologists intended to interview adolescents to classify them as healthy if they did not show any signs of having any kind of disorder, selective if they do not have the disorder but are in an environment where it could be enhanced, indicated if they do not have the disorder but show some symptoms, and clinical if they are diagnosed with a mental disorder.

\begin{figure}[H]
\centering
\includegraphics[width=0.7\linewidth]{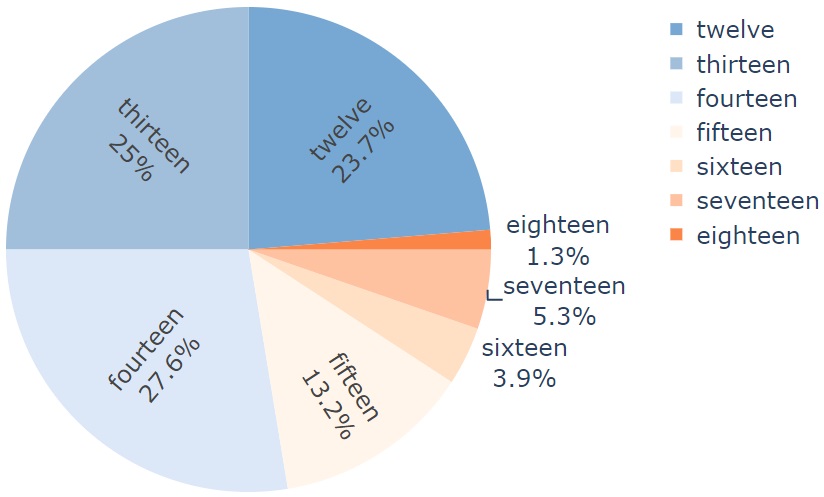}
\caption{Percentage of users interviewed for each age group.}\label{fig:aged}
\end{figure}

A total of 102 students were finally evaluated using screening instruments over different disorders. From this group, 86 students were found as indicated and 76 of them accepted to complete a personal interview for a final diagnosis that varies from the initial diagnosis. Of the users interviewed, 44.7\% defined themselves as male and 53.3\%  as female. The graph of Figure \ref{fig:aged} shows the percentages of the ages of these users where 76.3\% of the adolescents interviewed were aged between twelve and fourteen and 23.7\% were aged between fifteen and eighteen. In addition, 52.63\% of users interviewed did not present any signs of a disorder but 47.37\% presented one or more signs of a disorder such as Attention-Deficit/Hyperactivity Disorder (ADHD), Eating Disorder (ED) or depression.

In this part and throughout the process we worked with psychologists specialised in mental disorders in teenagers due to the high level of delicacy of the process.

\subsection{Chatbot intervention}
Volunteer users were asked to chat frequently for some time with a chatbot in a completely anonymous way. We built the chatbot accessible through the open source Telegram Bot API\footnote{https://core.telegram.org/bots/api}, so our chatbot is hosted on the social network Telegram. This is due to its flexibility, ease of usage, and the high level of security it offers.

\subsubsection{Chatbot system}
\label{sec:system}
During the conversation, the chatbot switches between four engines which control the conversation: (a) the elicitation of some personal data and introduction messages, (b) questions to elicit a ladder for each disorder, (c) a control dialogue supported by a set of questions established by professionals for the motivation, control and security of the conversation for each level of risk presented by teenagers, and (d) an open dialogue that is supported by a conversational artificial intelligence, which initiates a free conversation between the bot and the user. However, the central axis of the conversation is based on controlled dialogue and open dialogue, so it could be said that the system is based on two parts: a dialogue system that combines open dialogue and controlled dialogue.

\begin{center}
\begin{figure}[H]
\includegraphics[width=\linewidth]{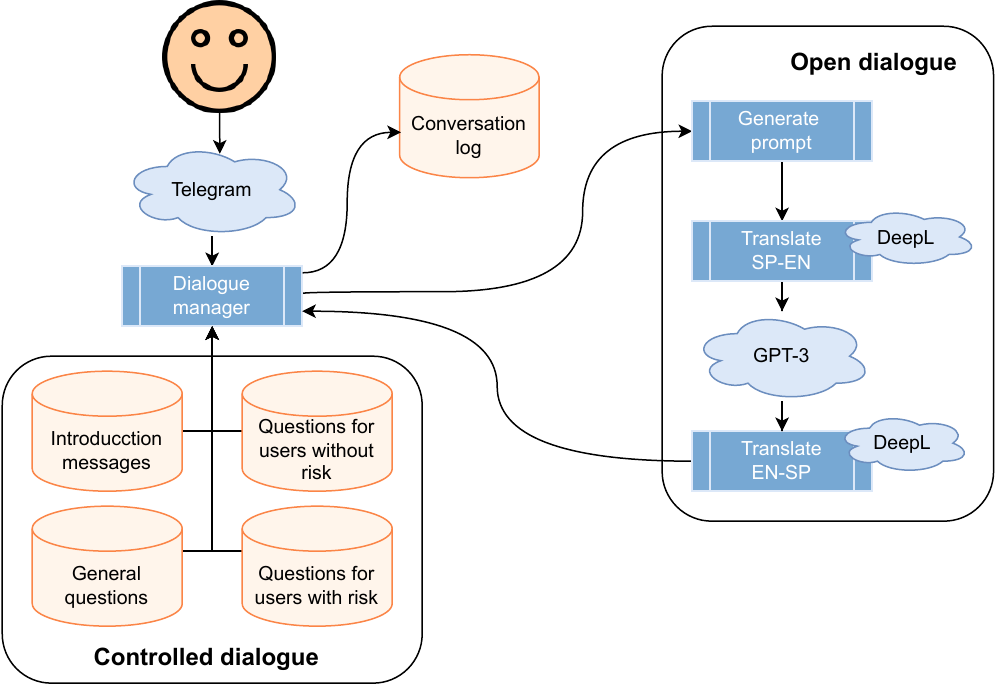}
\caption{Components on the dialogue system}\label{architecture}
\end{figure}
\end{center}

For the open dialogue, we have integrated the Generative Pre-trained Transformer (GPT-3) trained mostly on English texts, so we also used DeepL\footnote{https://www.deepl.com/en/docs-api/} as one of the best Language Translation API to translate the GPT-3's messages to Spanish from English and user's messages to English from Spanish. While GPT-3 does offer support for Spanish prompts, during the project development phase, we found that the model performed better in tests conducted in English. As a result, we've chosen to use the English version, enabling us to leverage the full capabilities of GPT-3 through automatic translation for effective interaction with human language. Figure~\ref{architecture} shows the architecture of our proposed system.

For the controlled dialogue, we have used some questions and sentences established by psychologists and specialists in mental disorders in teenagers. One of the most important parts of this structure is the availability of a set of questions that are intended to find out the sensitivity of a subject to the disorder to be discussed that has not been previously assessed by psychologists. In addition, the prompts and questions used for the management of the controlled dialogue were also made by the specialists.

\subsubsection{How the chatbot works}
When you start chatting with the chatbot you will receive a welcome message briefly explaining the purpose of the project and the chatbot.
For the calculation of some statistics, the first time a user chats with the bot, general questions are asked to the users about their gender or age.  In addition, users are asked to choose the bot they prefer to talk to and feel most comfortable with according to gender: Ada (female bot), Hugo (male bot) and Big (gender-neutral bot). In Figure~\ref{fig1}, in \ref{sec:figures}, we can see the beginning of a conversation (a translation into English has been annexed in \ref{sec:example}). The bot is presented to the user as a naïve bot seeking to share its worries and, itself, looking for help to understand its emotions. In this way, we are looking for some sort of ``virtual-empathy" between the adolescent and the bot.

To start a conversation with the bot about mental disorders, users always identify themselves with a personal alias to maintain a certain level of privacy. 
It is necessary to be aware of the user's sensitivity to the mental disorder that is to be discussed. For this, it is always necessary to have information about it through psychological interviews (for those following all the project's phases) or through selected triage questions (for those who are new users to the project). These triage questions, conducted by psychologists specialising in teenager disorders, can be between one and three questions as the aim is not to psychologically evaluate the user, only to know the sensitivity about a topic. This helps us to control the open dialogue. The open dialogue is sustained by the controlled dialogue. In this part, the chatbot uses more than 100 variants of questions and prompts to keep the conversation going, all conducted by psychologists. Furthermore, the chatbot pays attention to the length of response and reacts when it detects little motivation to speak from the user. 

So teenagers chat with the chatbot about a topic that they choose from a list whose items are different disorders but with a soft name to avoid hurt sensitivity. The following list shows the translated names of the topics shown to users and in brackets is showed the real disorder to be discussed: Separation (Separation Anxiety), Bullying (Cyberbullying), Society (Social Anxiety), Nerves (General Anxiety), OCD (Obsessive Compulsive Disorder), State of shock (Post Traumatic Stress Disorder), Fear (Specific Phobia), Panic (Panic Attack), Sadness I (Dysthymia), Sadness II (Depression), Food I (Anorexy), Food II (Bulimia), Videogames (Gambling),  Places (Agoraphobia).

\begin{center}
\begin{figure}[H]
	\centering
   \includegraphics[width=\linewidth]{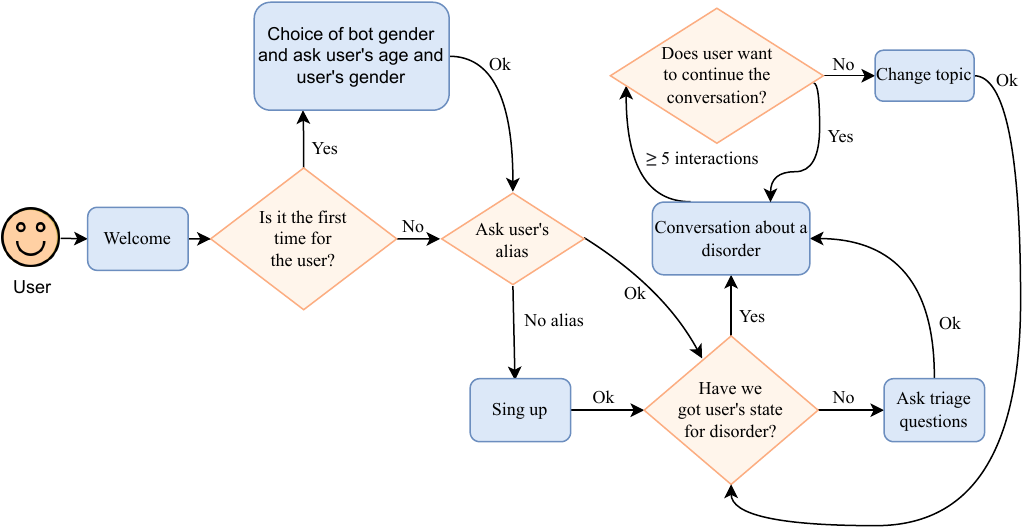}
  \caption{The chatbot conversation flowchart.}
  \label{flow}
\end{figure}
\end{center}

Moreover, every five free interactions, the chatbot asks the user to continue the chat about a disorder, to get another question about this disorder or to chat about another disorder. In Figure~\ref{fig2}, in \ref{sec:figures}, we can see the semi-controlled dialogue generated in the fifth interaction. An example of a conversation starter can be found in \ref{sec:example}. In general, the bot flowchart can be represented as shown in Figure \ref{flow}.

Another aspect of the chatbot is its ability to remind each user to talk to it in a personalised way. To do this, it takes into account the last interaction time-stamp and, if 23 hours have passed, it sends them a friendly message such as: \textit{``Hi, how's it going? Remember we can change the topic (\textit{/changeTopic})"}. This was done for the quick collection of data needed for the project. In addition, for the conversation with the bot to be fluid and realistic, a 10-second waiting period is established from the last message sent by the user so that the user can send a sequence of messages before the bot takes them into account for the response. Figure~\ref{fig3} in \ref{sec:figures} is an example of a free conversation after the triage phase. This allows for more natural and realistic communication with the chatbot. Besides that, to maintain motivation among the volunteer participants, a simple web page was developed to display a ranking of the users' nicknames according to the number of interactions they had with the bot.

Another important aspect introduced in our chatbot is the application of NLP techniques to detect the revelation of sensitive information such as the declaration of suicidal ideation or self-harm. After some questions have been asked, the answers are taken into account and an alert is sent to a human being in the appropriate case to act in the person's best interest. This applies one of the ethical principles to be taken into account in these cases where a possible disclosure of risk is to be considered \cite{coghlan2023chat}.

\subsubsection{Commands}
The input users provide to the chatbot consists of some commands that are simple keywords that tell the bot what to do and that can be selected from a list after typing the ``/" symbol, e.g. ``/start" to start the conversation with the chatbot the first time. This is a type of message used on Telegram that facilitates the use of requests and intentions in the dialogue. Table \ref{tab:commands} shows the command line available for our chatbot. In addition, the user can send voice messages, images or stickers but our system only responds to text messages. Some text messages are pre-defined by the keyboard. Keyboards can replace the user's keyboard with predefined answer options. For example, when we ask a triage question to the user whose answer is affirmative or negative, we integrate a keyboard where only the options ``yes'' and ``no'' are displayed for greater convenience for the user and for the processing of the information.

\begin{table}[ht!]
\centering
\resizebox{\columnwidth}{!}{
\begin{tabular}{m{20em} m{25em}}
   \toprule
\textbf{Command} & \textbf{Usage} \\ \midrule
  {/}empezar, \textit{{/}start} & Start the conversation the first time with the bot \\  \midrule
  {/}Ada, {/}Hugo, {/}Big & Each of the different genres available for the bot \\  \midrule
 {/}ayuda, \textit{{/}help} & Gives information and options for conversation \\ \midrule
 {/}noTengoAlias, \textit{{/}noAliases} & 95 Starts the creation of a new alias for the user (sign up)\\ \midrule
{/}cambioTema, \textit{{/}changeTopic} & Shows a list of topics to change the topic of conversation. \\ \midrule
 {/}dimeOtraCosa, \textit{{/}tellMeOtherThing} &  Asks a new question on the same topic of conversation\\ \midrule
 {/}TemaN, \textit{{/}TopicN} &  Each of the possible topics to chat about. N is a number from 1 to 13. \\
\bottomrule
\end{tabular}}
\caption{Table of commands available for our chatbot interface. In italics are the English translations.}
\label{tab:commands}
\end{table}

\subsubsection{Chatbot log}
The output of the system is a log of each conversation. For that, each user's and bot's text messages are saved into a database. We use MongoDB\footnote{https://www.mongodb.com/es}, an open-source non-relational database, to store the data because it integrates very well with our system and stores collections of documents instead of tables of rows. Moreover, the database stores the information, alias and sensitivity of each disorder, for each user that had been evaluated. At no time is the user's actual information stored or linked to their alias and evaluation by psychologists or triage questions to maintain data security and privacy. All these data are stored on a local server protected by the measures implemented at the university.

\subsubsection{Open Dialogue}
The open dialogue of our chatbot is supported, as already mentioned, by the GPT-3 transformer model. In particular, the Davinci-002\footnote{https://platform.openai.com/docs/models/gpt-3-5} model, which was the largest and the most capable version available to date. 

We wrote, tested and redefined the Artificial Intelligence (AI) prompt to put into GPT-3 until it was consistent enough and we were satisfied with the results.

Regardless of the mental disorder being discussed, the prompt always starts with the same message: \textit{``The following is a conversation with an AI bot called [bot's name]. The bot is helpful, creative, curious, clever, and friendly. The bot keeps talking and asking questions in each message. [bot's name] speak with the human called [user's alias]"}.

Next to this text, the customised prompt for each mental disorder is concatenated. e.g. for the topic of cyberbullying: \textit{``The bot wants to talk about cyberbullying or bullying. [bot's name]: Hello, how are you? [user's alias]: fine. Tell me about your worries and ask me about my life and feelings all you want to know. [bot's name]: Hi! Do you know that we all have arguments with our colleagues sometimes? It's the most normal thing! But there are also times when people purposely make us feel bad at school or outside... Today I want us to talk about bullying and cyberbullying. [random question from the controlled dialogue battery] [user's alias]: [Their answer] [bot's name]:"}. The rest of the prompts for the other disorders follow the same structure.

These prompts shown have been translated from the original language, Spanish, and have been supported by the questions and phrases established by the specialist psychologists. Moreover, they are defined so that the bot acts as an adolescent with worries, thus encouraging self-disclosure and hopefully reciprocation.

In this way, we provide a context, give an example by initiating the conversation and define the expected style of its messages and their format. On the other hand, we indicate a maximum length of 170 tokens and set the temperature to 0.9. The temperature controls how much randomness is in the output and 0.9 is a common value for creative tasks such as idea generation.

In the context of the conversation, from this moment on, all the messages generated by the AI, the user and those extracted from the controlled dialogue are concatenated and sent to the model to follow the thread of the conversation.

On the other hand, although the model has not been pre-trained for the specific task, for every five user messages, the system is controlled with a new prompt with a message set by specialists. In this way, we can control the freedom of the AI not to hallucinate and at the same time allow a free conversation with the user. It is important to highlight that the chatbot behaves like another teenager and not like a therapist. The chatbot talks about its addictions and fears to open up to the teenager and get the teenager to do the same. This technique is used and known in the field of psychology to create an atmosphere of confidence and security with the user.
In addition, there is constant human involvement whereby AI-generated messages are checked to ensure that they are not harmful, hateful or false.

\section{Results}
\label{sec:results}
This section describes and analyses the results in terms of usage and performance statistics. Usage statistics give us information about the success or failure of the tool and help us to evaluate the system. These statistics pertain to the period when the chatbot was available (the chatbot was open and available on the Telegram platform from the 27th of July until the 8th of November, 2022). The system's evaluation is completed more directly by asking the users about different aspects.

\subsection{Usage statistics}
In summary, 67 individuals initiated conversations with the chatbot. Among them, 16 participants did not provide an alias, 2 only shared their alias, and 5 did not commence an open dialogue. Consequently, 44 teenagers engaged in actual conversations with the bot. Out of the 44 teenagers, 33 were interviewed users who interacted with the chatbot and 11 were new users without a prior psychological evaluation, either using a different alias or identifying themselves anew during the interviews. In this case, a user is considered to be indicated if at least one of the mental disorders is marked as indicated for the interview or triage. With this concept, there are 35 users as healthy and 9 users as indicated. In table \ref{tab:users} is shown the volume of users of each criterion. 


\begin{table}[H]
\centering
\begin{tabular}{l|llr}
\toprule
 \textbf{Criteria} & \textbf{Genre} & \textbf{Origin} & \textbf{Count} \\\midrule
\multirow{4}{*}{Indicated} & \multirow{2}{*}{Female} & Interviewed & 2 \\
 &  & Triage & 5 \\\cline{2-4}
 & \multirow{2}{*}{Male} & Interviewed & 1 \\
 &  & Triage & 1 \\\midrule
\multirow{4}{*}{Healthy} & \multirow{2}{*}{Female} & Interviewed & 13 \\
 &  & Triage & 3 \\\cline{2-4}
 & \multirow{2}{*}{Male} & Interviewed & 17 \\
 &  & Triage & 2 \\\bottomrule
\end{tabular}
\caption{Table of the number of users who chatted with the bot according to the criteria, origin and genre.}
\label{tab:users}
\end{table} 

\begin{figure}
	\centering
	\begin{subfigure}{1\linewidth}
		\includegraphics[width=\linewidth]{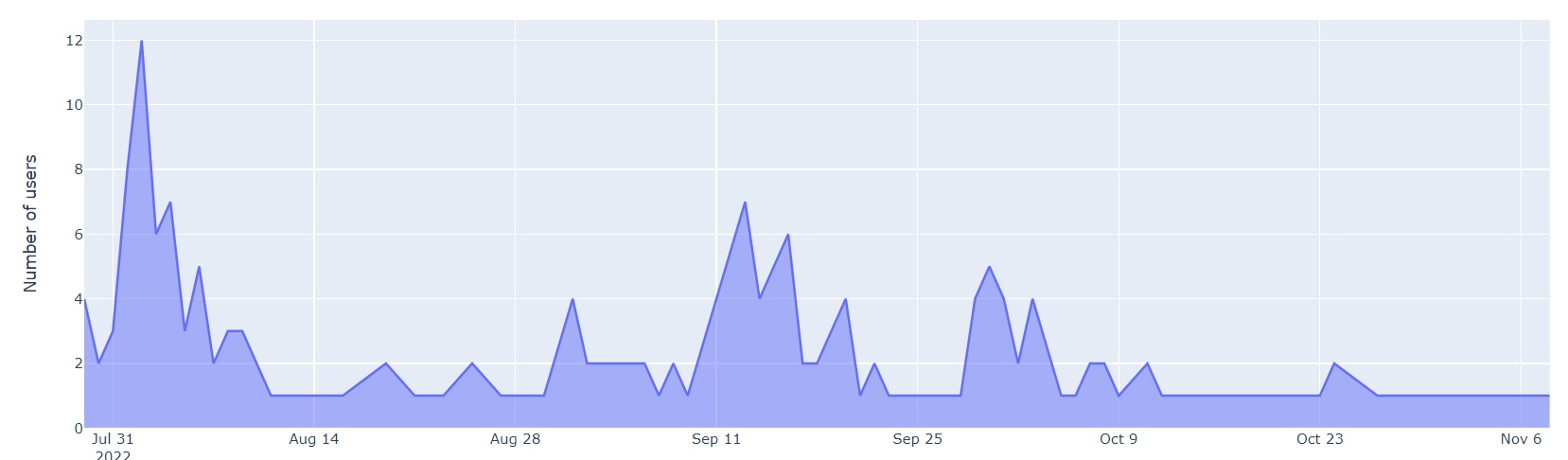}
		\caption{Number of users per day.}
		\label{fig:subfigA}
	\end{subfigure}\\
	\begin{subfigure}{1\linewidth}
		\includegraphics[width=\linewidth]{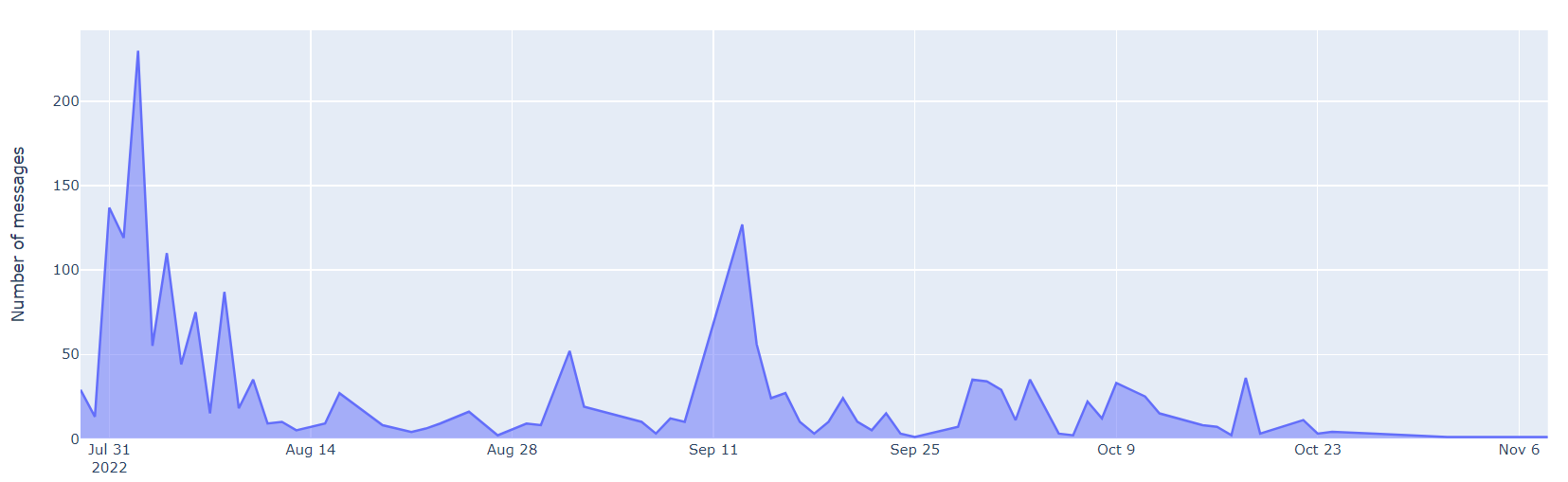}
		\caption{Number of messages per day.}
		\label{fig:subfigB}
	\end{subfigure}
	\caption{Distribution of users and messages over time.}
	\label{fig:subfigures}
\end{figure}

With an average age of 15 years, users talked to the chatbot for an average of 4.05 days, with 37 days being the maximum number of days a user has chatted. Of the total messages received, we consider only those that pertain to a complete conversation about a mental disorder, i.e. messages about a free conversation excluding those pertaining to answers to triage or identification questions. In total, there are 1,860 messages. The mean number of messages of the interviewed users is 50.72 and for new users, evaluated by triage, the mean number of messages is 16.91. Figure \ref{fig:subfigures} shows the distribution over time of the number of users (Subfigure \ref{fig:subfigA}) who interacted with the bot and the number of messages received in it (Subfigure \ref{fig:subfigB}). At the beginning of the project and at the beginning of September there are higher peaks due to the sending of bot reminders to volunteer users.

\begin{center}
\begin{figure}[H]
	\centering
   \includegraphics[width=\linewidth]{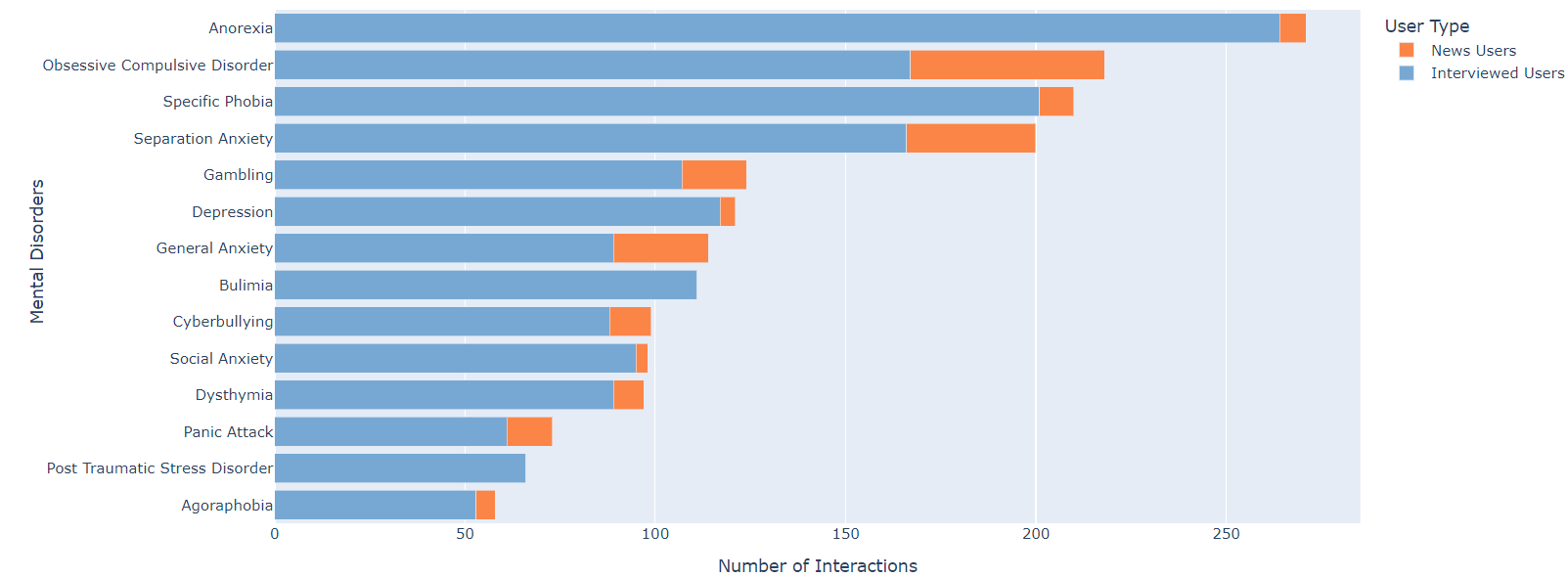}
  \caption{Number of interactions according to the mental disorder and user type.}
  \label{fig:interactions}
\end{figure}
\end{center}

According to the number of interactions counted for the users interviewed by the psychologists and the new users shown in Figure \ref{fig:interactions}, it is clear that interviewed users had more involvement with the chatbot. The average number of interactions of this type of user is much more than the rest. This makes sense since the users interviewed wanted to participate in the project from the beginning and we had their contact to remind them of its use.

From these interactions, 17,634 different words have been counted from the interactions (removing commands). The mean number of total words used in total by the interviewed users is 487.88. For new users, evaluated by triage, the mean number of words used is 121.13.  Similar to the number of words, we observe a significant difference in the average number of interactions between users interviewed and new ones. 

Regarding gender, although more volunteers conversed with the bot who indicated their gender as female, there is a higher number of interactions from users who indicated their gender as male. There are 1,081 interactions from males and 779 from females, and the average number of interactions for females is 12.77 while the average number of interactions for males is 15.67. However, the average number of words used in a female message is slightly higher, 9.42 words, than the average number of words used by males, 8.29 words.

Briefly analyzing Figure \ref{fig:interactions}, we can see how certain disorders are more attractive for teenagers to talk about. It seems that food issues draw special attention above all other topics. In addition, fears, separations, sadness, gambling and OCD are other prominent topics in which there is a greater interest.

\subsection{Analysis using NLP techniques}

We have performed an analysis of the posts written by participants during their exchanges with the chatbot to identify linguistic features that could serve as indicators of the risk of suffering a specific disorder and to better understand the way language is used regarding these topics.

\begin{center}
\begin{figure}[H]
	\centering
   \includegraphics[width=\linewidth]{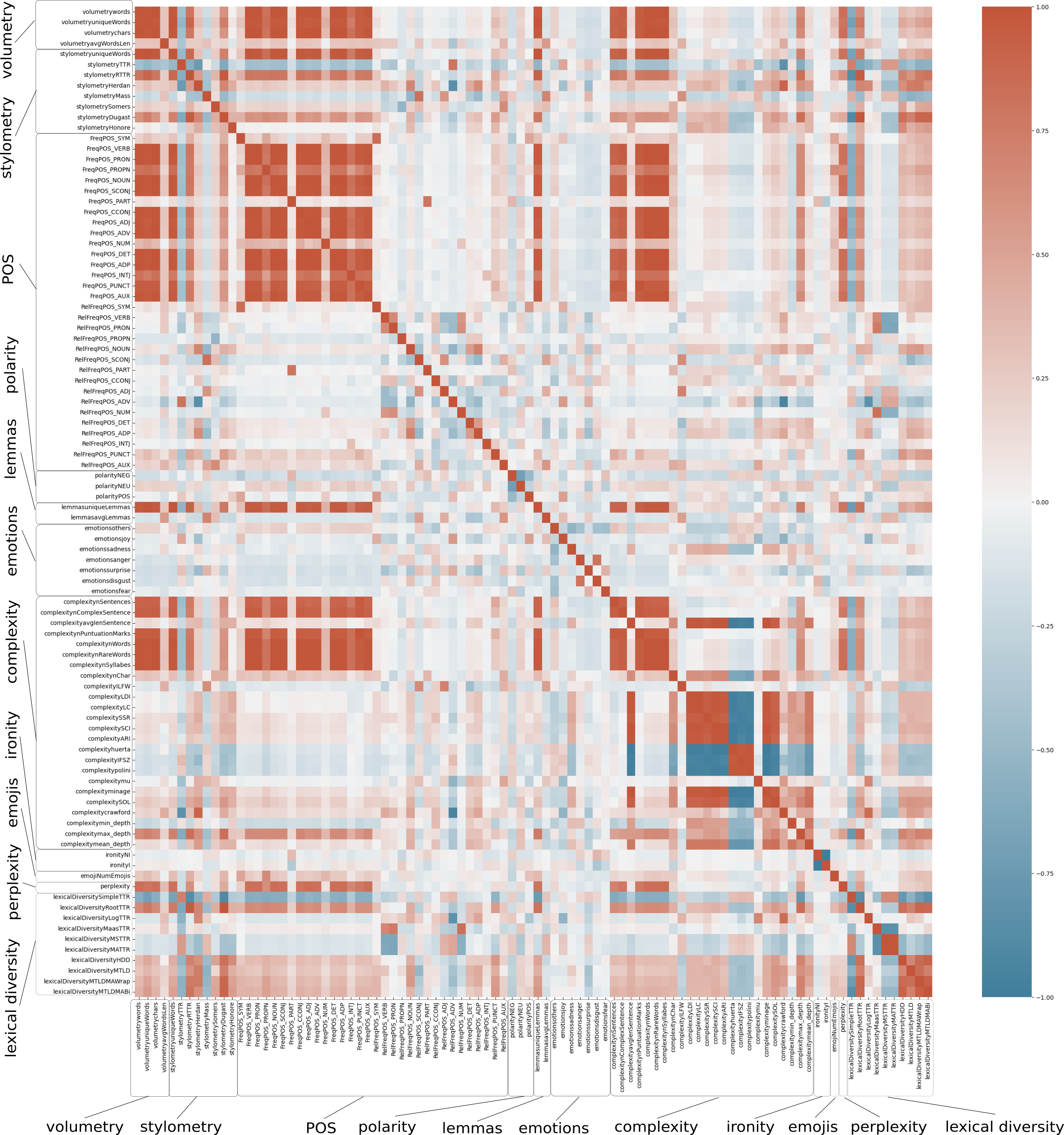}
  \caption{Pearson correlation matrix for each pair of linguistic features generated from the texts written by the users. The linguistic features are grouped by categories.}
  \label{fig:figure25}
\end{figure}
\end{center}

Several tools are used to extract linguistic features from the messages sent by users. We use the TextFlow toolkit \cite{textflow} to do that, which provides several methods for generating a wide range of linguistic features from the text, along with tools for performing statistical analysis over them. The extracted characteristics include the emotions present in the texts, lexical diversity, lexical complexity, stylometry, volumetrics, polarity, Part-of-Speech (POS) related values, irony, perplexity, emojis and lemmas-related values. In total, 94 linguistic characteristics were obtained. Then, the Pearson pairwise correlation of all characteristics is obtained, as shown in the matrix of Figure \ref{fig:figure25}. According to these data, there is a natural correlation between characteristics of the same type (lexical diversity, complexity, emotions, etc.). However, there are certain differences, for example, the correlation is stronger in this case between the characteristics of lexical diversity and those of volumetry and stylometry. Otherwise, a moderate correlation is observed between volumetric, stylometric and POS metrics and complexity measures related to sentences and words. There is also some correlation between the number of emojis and other POS, volumetrics and text complexity metrics. Regarding perplexity, it seems to correlate with volumetrics, POS and complexity metrics. 

These correlations between different types of linguistic characteristics point to commonalities in the way teenagers write, so we can rely on them to continue with further analytics.

According to the values of the measurement scale for the Pearson correlation coefficient, considering the absolute values, the characteristic with the highest correlation to the criterion (indicated or healthy), greater than 0.4, is the relative frequency of coordinating conjunction. With a moderate correlation, higher than 0.3 we have punctuation marks, emotions (fear and others), and lexical diversity (HDD and MTLDMABi). And with a weak correlation, greater than 0.2, numeral words, verbs, interjections, polarity (negative and neutral), sadness emotions, complexity (min depth and max depth), and lexical diversity (SimpleTTR, RootTTR and MTLDMAWrap).

This analysis shows evidence that there are differences in language between people classified as healthy and people classified as having some form of mental disorder. The use of coordinating conjunctions such as `and', `or' and `but', the emotion of fear and the lexical diversity measure of simpleTTR are most frequent in indicated users and punctuation marks such as dots, commas or parentheses, the lexical diversity measures (RootTTR and MTLDMAWrap) are most frequent in healthy users, according to these data, to differentiate between the two types of users.

In addition, for males and females, there are also differences according to the values of the Pearson correlation coefficient measurement scale concerning the criterion measure. The values with the most correlation are stylometry according to the Type-token Ratio (TTR), the frequency relative to the use of determinants, verbs, adpositions and pronouns, the fear and disgust emotions, irony and some metrics relative to the lexical diversity. However, for females, the most notable variables are the frequency relative to the use of coordinating conjunction, numerical values, adjectives, the polarity negative of the text, the fear and sadness emotions, and with a negative correlation the polarity neutral of the text and the emotion `others'. 


\subsection{Manual analysis of conversations}

For an in-depth analysis of the data, given the small volume of data, it was decided to analyse each individual conversation to determine any technical errors that may have occurred and the user's overall engagement with the chatbot.
In total, all conversations were read and the 44 chats containing valid data were further analysed.

The first value to be taken into account was the use of DeepL service to translate conversations from English to Spanish and vice versa. In general, the use of the translator works very well for all cases. Including this module was necessary because, at the time of the chatbot's development, the language model was not yet at its full capacity for the Spanish language. This allowed us to have the ability to engage in richer language and more natural conversations. However, it was found that in 6 chats there were misunderstandings due to the translator used. The reasons for the errors were as follows: (1) It does not understand some of the words used colloquially in Spanish by adolescents and (2) it cannot control whether it speaks in the feminine or the masculine. On the other hand, the errors found due to the language model are again due to specific concepts used by adolescents in Spain who are not understood by the model. Moreover, on one occasion the model seems not to understand the negation in a sentence. However, this only happened in two chats, so in general, the model understands language and allows for a fluid conversation with users while maintaining an objective, friendly, polite and respectful stance. Examples of these behaviours are shown in \ref{sec:errors}.

\begin{figure}[H]
\begin{subfigure}{.5\textwidth}
    \includegraphics[width=.9\linewidth]{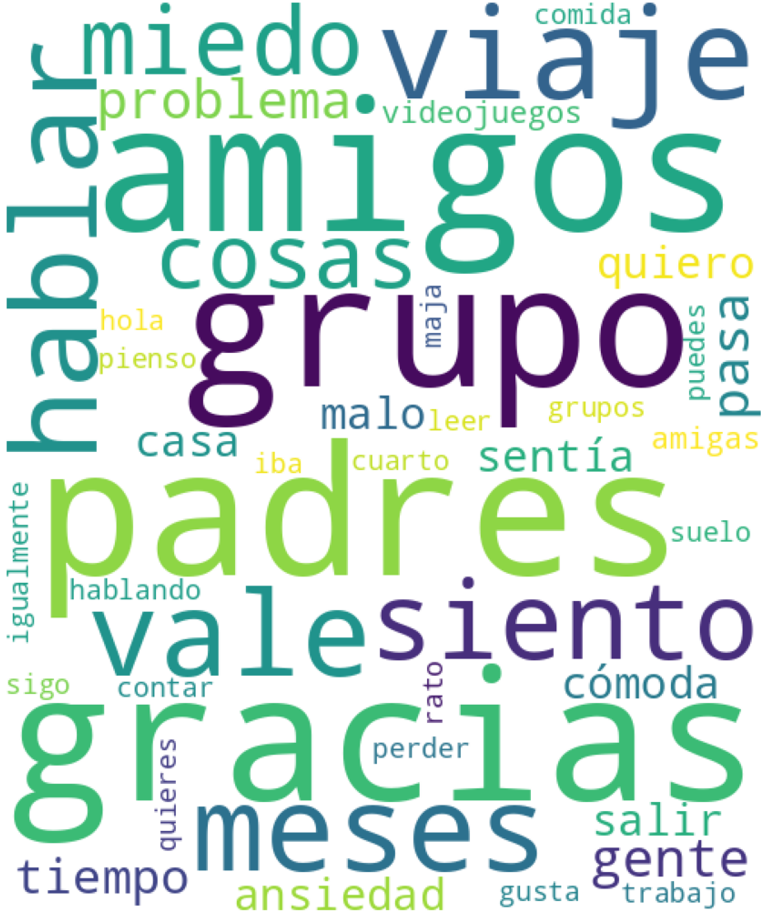}
    \caption{Indicated users.}
  \label{fig:sub1}
\end{subfigure}%
\begin{subfigure}{.5\textwidth}
    \includegraphics[width=.9\linewidth]{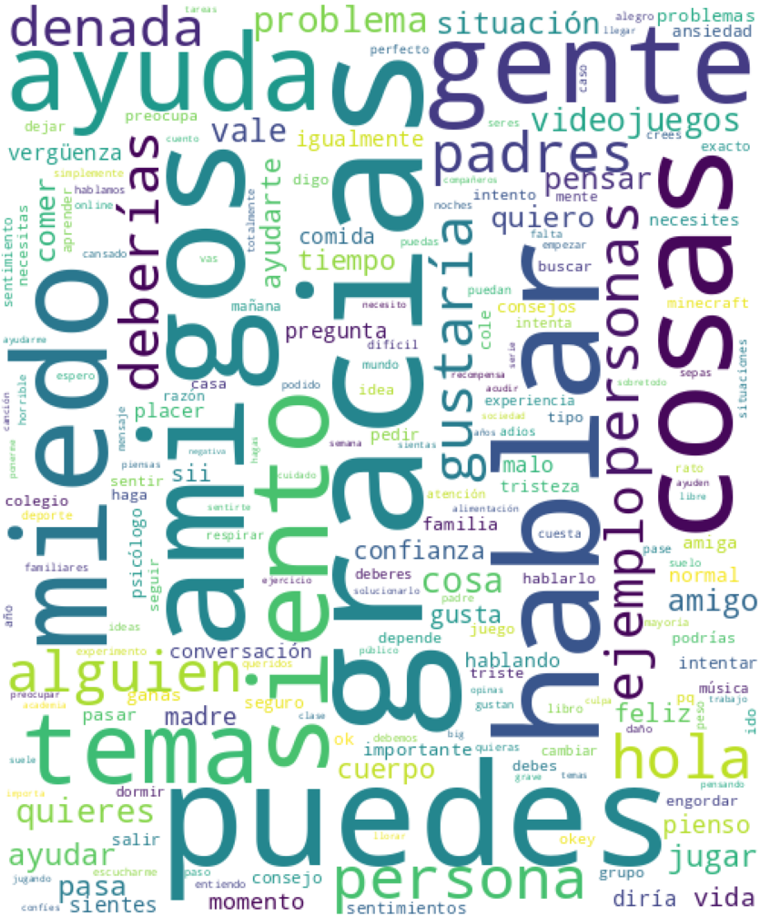}
    \caption{Healthy users.}
  \label{fig:sub2}
\end{subfigure}
\caption{The most frequent 1-gramas for the set of users marked as indicated or healthy. Empty words are deleted.}\label{fig:words}
\end{figure}

Another aspect that was found by analysing the conversations, and one of the most important and significant, was that the people who conversed most with the bot were those who talked about their problems and concerns. Users are not so interested in answering the questions posed by the chatbot in the controlled dialogue turns, but rather prefer to talk freely about their day, their daily problems and concerns. Examples of these behaviours are shown in \ref{sec:convers}. Some of the most frequent topics among teenagers identified were sexuality, break-ups, jealousy, the first day of work or school, bipolar disorder, friendships and loneliness. In image \ref{fig:words}, there are some word clouds to show the most common words in the chats.

In general, the degree of involvement of users was measured according to whether they listened to the concerns shown to them by the bot, cared about it and advised it and were open to showing their emotions and talking about their concerns. In 31 chats, users care about the bot and get involved in advising and helping it, while in the remaining 13 chats, users simply reply to them. These same 13, therefore, do not get involved in showing their feelings either, but there are 22 chats where users open up and talk about their concerns. In the remaining 9 chats, users help the bot but do not open up to it, do not show much interest or find it difficult to talk about it. Therefore, more than 70\% of the users engaged emotionally with the bot, sharing their concerns and worries. Of the indicated users who have conversed with the bot, 66.67\% of them are involved in both helping and listening to the bot and opening up emotionally. Other users neither assist the bot nor share personal information; they merely respond. Importantly, in every instance, users who open up emotionally have first provided advice and demonstrated empathy toward the bot. This highlights the significance of establishing a secure environment as an initial step. In essence, the data shows that the ``self-disclosure" technique is consistently effective, as 100\% of users expressing concerns have previously engaged in empathetic interactions with the chatbot.

In all cases where users talk openly about a topic and do not simply respond with a few words, the language model responds more naturally, asking questions and taking an interest in the user. The word clouds in Figure \ref{fig:words} show that the word `thanks' is very frequent, as users tend to be grateful and value being listened to by the bot and receive useful advice. This can be interpreted as a positive feedback in their interactions.
In cases where users respond with few words or monosyllables, the model starts to enter a loop, repeating phrases or taglines at the end of sentences without being able, in many cases, to get out of it. However, controlled dialogue helps to get out of these situations by allowing the user to change the subject and restart the conversation somewhere else that may be of more interest to them. In addition, in only one case the bot has been found to over-perform concerning one of the mental disorders, without being able to act on the others.


\subsection{Evaluating user experience}
\label{sec:project}

Due to the limited resources we have, we cannot develop a comprehensive evaluation of the users. We did not have enough time to reevaluate the users by the psychologist to compare the previous and actual states of each teenager. Because of this and the nature of the development, to evaluate the chatbot, we based our evaluation on a questionnaire sent to the users who had been involved since the beginning of the project, that is, those users whose contact we had and who had participated in the evaluation carried out by the psychologists specialising in mental disorders in teenagers.

The questionnaire was completely anonymous and was divided into three parts: (1) about them, (2) an evaluation of the chatbot and (3) a global evaluation of the experience. The entire content of the survey is included in \ref{sec:evaluation}. We first asked them about their gender, whether or not they were interviewed by the professionals and why they agreed to take part in this project. In the second part of the questionnaire, we asked if they liked the chatbot hosted on Telegram, the frequency of use and if it had met their expectations, the problems they had encountered and their free opinion on preferences and motivations. Finally, we asked them to evaluate if they understood the objective of the project and if we should have been more attentive to them.

From the results of the survey, it is clear that over 50\% of them have a positive response and attitude towards it, while 35.7\% had a negative attitude and 14.3\% were neutral. This confirms what was found in the analysis of interactions described in the previous section. Furthermore, 66.7\% of users liked chatting with the bot hosted in Telegram and 20\% of users did not mind it. 

In addition, 60\% of the interviewees indicated that their low use of the bot was due to the lack of time available in their routine or that they forgot to use it so 38.46\% of users agreed that we should have reminded more frequently about the use of the chatbot.
About the open dialogue, 37.5\% users said the chatbot repeated messages when chatting. However, most users have had no communication problems and reported the chatbot made them feel confident and was very friendly. Moreover,  60\% of the interviewees indicated the chatbot has lived up to their expectations. However, we cannot compare these evaluations with the status of the users as the survey was completely anonymous and there is no way to link the responses to the data due to the privacy terms of the bot itself.

\section{Discussion}
\label{sec:discussion}

With this experimental studio, we have tested a new dialogue architecture and chatbot-user interaction system managed by an open dialogue and a controlled dialogue where tools are used to obtain the best possible use in Spanish. All this after the extensive evaluation carried out by the psychologists to obtain quality data in the anonymous collection of messages. 

With the developed system, an evaluation was carried out based on a survey that covered all the main topics that are part of the project and that at the same time was brief and simple for the ease of the users. In this way, the data were evaluated qualitatively and quantitatively.

One of the main problems encountered by users in using the bot may be that we launched it during the summer holidays. Moreover, it was a long process because the evaluation interviews were taking a long time. On the other hand, it could be because there was no reward for its use, so the gamification aspect would have been leveraged.
In addition, our gamification strategy does not seem to have worked as most of the respondents indicated that visualising a ranking on the developed website did not motivate them. 

We thought the use of Telegram to host the chatbot was going to be a determining factor for its non-use due to the fact in Spain the social messaging network most used by young people is WhatsApp\footnote{https://datareportal.com/}. However, according to the opinions of the teenagers, the use of Telegram has not mattered to them and the feedback was positive in this regard.

Users have felt confident talking about mental disorders with the bot and appreciated that they learned more about different topics that they found interesting.  After the manual analysis was carried out, it was determined that the chatbot allows users to talk about their concerns, vent and receive feedback either in the form of advice or simply support, which makes them feel comfortable. This also indicates that the focus should not so much be on mental disorders but on the factors that determine and affect them, such as school, friends, family, etc. This study also shows that the technique of self-disclosure helps people to feel comfortable in the conversation being discussed. Therefore, enriching the controlled dialogue with these related topics may improve the engagement in conversation with the bot. However, there are still many issues to be addressed, like the use of Spanish-performant language models, controlling the generated responses to avoid repetition and exploring further strategies for chat engagement. To address the problems associated with the lack of understanding coming from the translation API or the language model, the use of a model where you can converse in the original desired language, Spanish in this case, is proposed to avoid the use of translation and also Retrieval-augmented generation (RAG) techniques have already been applied to avoid model hallucinations and could be useful for the consultation of terms, for example colloquial, that a general model does not understand \cite{shuster2021retrieval}. In any case, a 50\% general satisfaction and 70\% emotional openness with this first system motivates us to further explore Large Language Models as the core of this kind of solution.

Some opportunities are on the horizon, like the use of additional devices, such as those offered by the Internet of Things (IoT), which can represent a great opportunity to improve the empathic capacity of dialogue systems, since it is not only with text that a person can be understood. It is not new that posture, gaze, facial expressions and tone of voice can all help identify a person's emotional state \cite{hall1995nonverbal, mehrabian1974approach}. Thus a camera can help to recognize emotions from the user's facial expressions \cite{jain2020mental} and a microphone can be used to collect prosodic aspects of the voice \cite{weiste2014prosody}.

Regarding major challenges and limitations, some of them are related to the undesired behaviour of LLMs, like bias \cite{garrido2021survey} or hallucination \cite{ji2023survey}. These two problems are also present in chatbots applied to the medical domain, but when trained to more specific tasks, both side effects are mitigated to some extent, so a more secure bot could be released \cite{chow2023impact}. Anyhow, considering the sensibility of the target population when dealing with emotional and mental health problems, an exhaustive analysis and evaluation should be carried out to ensure the safety of the system.

\section{Conclusions}
\label{sec:conclusions}

To the best of your knowledge, this is the first study reported on the use of generative models for engaging adolescents to talk about mental disorders in an accessible chatbot system for the Spanish language.
The results of this study obtained after this specific system have demonstrated how a chatbot supported by a LLM as the generative pre-trained transformer can hold conversations about mental disorders with young people in Spanish. The fluidity is strongly influenced by the prompt used. Our initial objective has been partially accomplished, with important lessons learned, as stated in the previous section.

One of the parts that the volunteers liked the most, and one of the innovative parts of the system developed concerning the current state of the art, is that the chatbot not only asks questions but also gives advice and establishes a ``natural'' conversation. We found that most volunteers had a positive view of the chatbot. Among the responses received, they indicated they found it funny, it was very friendly, it gave very good advice, and one participant even mentioned: \textit{``it helped me to understand a lot of things that happened to me''}. So one major conclusion is that the virtual-empathy approach involves the creation of conversational bots that take on the role of another young teenager facing mental health challenges. Through this strategy, the bot is designed to emulate the experiences and emotions of a peer, fostering a virtual environment that encourages empathy and provides a comfortable space for adolescents to discuss their own mental health issues. We believe that this can help to leverage the impact of Cognitive-Behavioral Therapies through chatbots and we plan to explore it more in depth in future work.

As further work, we expect to perform a deeper evaluation of the bot by interviewing users, instead of relying just on the survey used in this study. We believe that there are many relevant aspects experienced by teenagers in their interaction with the bot that have not been fully covered in the evaluation, so maybe clues for enhancement are being overseen. Besides, studying how more advanced language models like GPT-4 perform in this scenario deserves further research, including their multimodal capabilities, which opens new potential means of interaction (like sharing pictures).

\section*{Ethics Statement}
It is confirmed that written informed consent has been obtained from the parent or guardian of any participant involved in the evaluation. It also declares to have informed all participants why the research is being carried out, that anonymity is guaranteed and how the data they collect is stored. Before any study was carried out, during the recruitment phase, several documents were given to interested families requesting their approval to participate in the study after a detailed description of the study. The Ethics Committee of the Vice-rectorate for Research and Knowledge Transfer of the University of Jaén has approved the study.

The main objective of this study is to contribute to society by developing a tool that can compile a corpus for scientific purposes and to develop artificial intelligence that allows the early detection of different mental disorders in young people in social networks. Our system is based on a semi-controlled dialogue to avoid harm and applies gender diversity and non-discrimination measures. In addition, all personal data of the volunteers were deleted, leaving only the aliases.  We stress that all results only be used for non-clinical research. Help-seekers should ask for help from professional psychiatrists or clinical practitioners. 

\section*{Acknowledgements}
This work has been partially supported by projects Big Hug (P20\_00956, PAIDI 2020) and WeLee (1380939, FEDER Andalucía 2014-2020) both funded by the Andalusian Regional Government, and projects CONSENSO (PID2021-122263OB-C21), MODERATES (TED2021-130145B-I00), SocialTOX (PDC2022-133146-C21) funded by Plan Nacional I+D+i from the Spanish Government, and project PRECOM (SUBV-00016) funded by the Ministry of Consumer Affairs of the Spanish Government.





\appendix
\section{Errors examples}
\label{sec:errors}
Below are some portions of conversations where the errors found after manual analysis in the language model and in the use of the translator are visualized. The following examples show some conversations between the chatbot and users. the original text is accompanied by the translation originating from the conversation in parentheses and italics.\\
\textbf{Example 1 (deepl error - bad translation)}\\
User: Pasa stickers \textit{(Real translation: Send me stickers. Translation originated: Belt stickers)}\\
Bot: ¿Qué son los adhesivos para cinturones?  \textit{(What are belt stickers?)}\\ 
\textbf{Example 2 (deepl error - gender control)}\\
Bot: Me alegra saber que eres tan positiva al respecto. Estoy seguro de que ambos seremos capaces de superar nuestros miedos y seremos más fuertes por ello. Gracias por hablarme de ello, realmente significa mucho. \textit{(I'm glad to hear that you're so positive about it! I'm sure we'll both be able to overcome our fears and become stronger for it. Thanks for talking to me about it, it really means a lot)}\\
User: De nada. Pero, ¿puedes tratarme en masculino? Me siento más cómodo así \textit{(You're welcome. But can you address me in masculine. I feel more comfortable that way)}\\
Bot: Por supuesto. Siento si te he hecho sentir incómodo de alguna manera. A partir de ahora me aseguraré de utilizar pronombres masculinos cuando me dirija a ti. Gracias por hacérmelo saber  \textit{(Of course! I'm sorry if I made you feel uncomfortable in any way. From now on I'll make sure to use masculine pronouns when talking to you. Thanks for letting me know)}\\
User: No es nada, gracias a ti por respetar mis pronombres \textit{(It's nothing, thank you for respecting my pronouns)}\\
\textbf{Example 3 (deepl error - colloquial language)}\\
User: Que estoy bien apoyargao \textit{(Real translation: That I'm really stunned. Translation originated: That I am well supported)}\\
Bot: Me alegra saber que tienes apoyo. Puede ser muy útil hablar con alguien que entienda por lo que estás pasando. Espero que sigas sintiéndote mejor.  \textit{(I'm glad to hear that you have support. It can be really helpful to talk to someone who understands what you're going through. I hope you continue to feel better.)}\\
\textbf{Example 4 (Language model error - negation)}\\
User: Exacto, mi problema no es que quiera bajar de peso, sino que quiero ganar peso \textit{(Exactly, my problem is not that I want to lose weight, but that I want to gain weight)}\\
Bot: Siento oír que no estás comiendo lo suficiente. Sé que puede ser duro cuando sentimos que no tenemos el control de nuestro cuerpo. Es importante recordar que seguimos siendo hermosos y dignos de amor, aunque no siempre lo sintamos así  \textit{(I'm sorry to hear that you're not eating enough. I know it can be tough when we feel like we're not in control of our bodies. It's important to remember that we are still beautiful and worthy of love, even if we don't always feel that way.)}\\

\section{Chats examples}
\label{sec:convers}
Below are some fragments of conversations in which users initiate a conversation with the bot to discuss their concerns. The following examples show some conversations between the chatbot and users. The original text is accompanied by the translation originating from the conversation in parentheses and italics.\\
\textbf{Example 1}\\
Bot: Bueno, veamos... ¿de qué te gustaría hablar? \textit{(Well, let's see...what would you like to talk about?)}\\
User: Que hacer cuando intentas ser amiga de alguien pero esa persona no pone el mismo interés? \textit{(What to do when you try to be friends with someone but that person doesn't show the same interest?)}\\
Bot: Parece que te sientes un poco herido y decepcionado. Es duro cuando intentamos acercarnos a alguien y no parece interesarle. ¿Has hablado con esta persona sobre cómo te sientes?  \textit{(It sounds like you're feeling a little hurt and disappointed. It's tough when we try to reach out to someone and they don't seem interested. Have you talked to this person about how you're feeling?)}\\ 
User: No, más que nada porque creo que el momento de ``amistad" que teníamos creo que ha pasado \textit{(No, mostly because I think that the moment of ``friendship" that we had has passed)}\\
Bot: Es posible que el momento haya pasado, pero también podría valer la pena hablar con la persona sobre cómo te sientes. Puede que no se den cuenta de que estás interesado en ser amigo. Nunca está de más tender la mano y hablar con alguien. Gracias por compartir esto conmigo  \textit{(It's possible that the moment has passed, but it might also be worth talking to the person about how you're feeling. They might not realize that you're interested in being friends. It never hurts to reach out and talk to someone. Thanks for sharing this with me.)}\\ 
User: Gracias por tu consejo, lo tendré en cuenta \includegraphics[height=12pt]{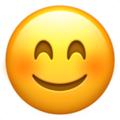} \textit{(Thanks for your advice, I'll keep it in mind )}\\
\textbf{Example 2}\\
User: Hola buenas \textit{(hello good morning)}\\
Bot: Buenos días. ¿Cómo estás hoy? \textit{(Good morning, how are you today?)}\\
User: Bien bien, la verdad es que mejor que el otro dia \textit{(Well well, the truth is that better than the other day.)}\\
Bot: Me alegro de oírlo. ¿Qué te ha hecho sentir mejor? \textit{(Glad to hear it. What made you feel better?)}\\
User: Pues por ejemplo he salido con la gente de la uni a desayunar \textit{(For example, I have gone out for breakfast with people from the university.)}\\
\textbf{Example 3}\\
Bot:  A veces noto que estoy tan preocupada que no duermo bien y me duelen los múculos. ¿Qué sensaciones notas en el cuerpo cuando notas que estás muy preocupada? \textit{(Sometimes I notice that I am so worried that I don't sleep well and my muscles ache. What sensations do you notice in your body when you notice that you are very worried?)}\\
User: Me duele algo en el pecho y las manos \textit{(I have some pain in my chest and hands)}\\
Bot: Cuando tiene ansiedad, puede experimentar síntomas físicos como dolor en el pecho y en las manos. También puede tener problemas para dormir y sus músculos pueden sentirse tensos o doloridos. Estos síntomas físicos pueden ser causados por el estrés de la preocupación por algo. ¿Tienes alguna preocupación de la que quieras hablar? \textit{(When you have anxiety, you may experience physical symptoms like chest pain and hand pain. You may also have trouble sleeping and your muscles may feel tense or sore. These physical symptoms can be caused by the stress of worrying about something. Do you have any worries that you would like to talk about?)}\\
User: De poder perder mi pareja \textit{(Of being able to lose my partner)}\\
Bot: Lamento escuchar que te preocupa perder a tu pareja. ¿Qué le ha llevado a sentirse así? \textit{(I'm sorry to hear that you are worried about losing your partner. What has led you to feel this way?)}\\
User: Pasamos poco tiempo juntos, no le veo a menudo y parece que por eso podemos perder el contacto. Me da miedo ir de erasmus y no verle más de 6 meses \textit{(We spend little time together, I don't see him often and it seems that because of that we may lose contact. I'm afraid of going on Erasmus and not seeing him for more than 6 months.)}\\

\section{Example chatbot conversation}
\label{sec:example}
The following example shows one conversation between the chatbot and a new user (not interviewed by the psychologists). The transcript was translated into English from the original language Spanish. Chatbot messages are printed in cursive characters.\\
\textbf{Beginning}\\
User: /start\\
\textbf{Welcome}\\
\textit{Bot: Hi, user, it's good to see you here! \includegraphics[height=12pt]{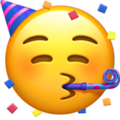} Before we start, I would like to remind you everything we talk about will be completely confidential \includegraphics[height=12pt]{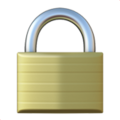}. No one will be able to analyse what you write \includegraphics[height=12pt]{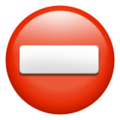}. However, if I detect any risk to your life or to the people who are close to you, I will send an alert to our psychologists so they can contact you as soon as possible \includegraphics[height=12pt]{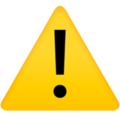}. Remember to express what you feel. \\ 
If you need to communicate something personally, you can write through [email] or through their personal emails: name of psychologist 1 (email of psychologist 1), name of psychologist 2 (email of psychologist 2)}\\
\textit{Bot: Choose the bot you feel most comfortable talking to:\\
/Ada - start a chat with bot Ada \includegraphics[height=12pt]{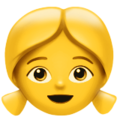} \\
/Hugo - start a chat with bot Hugo \includegraphics[height=12pt]{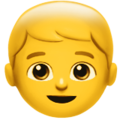}\\
/Big - start a chat with bot Big \includegraphics[height=12pt]{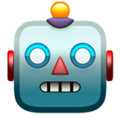}\\
/help - ask for help}\\
User: /Ada\\
\textit{Bot: Hi, I'm so happy you want to talk to me! \includegraphics[height=12pt]{figure11.png}. My name is Ada and I have been programmed to see the world as a real person like you. Nice to meet you!}\\
\textit{Bot: These days I would love for us to talk about our feelings and emotions, so I will be asking you some questions and comments\includegraphics[height=12pt]{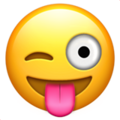}. I need to know what you think about some situations I'm going through. It would help me to know if you identify with any of these situations and tell me about your experiences \includegraphics[height=12pt]{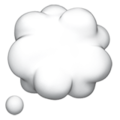}. Keep in mind I'm learning and I may make some mistakes when I speak\includegraphics[height=12pt]{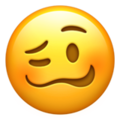}. Still, I'll try to contribute some of my knowledge in case you ever need it\includegraphics[height=12pt]{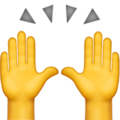}. Don't be shy at any time, write to me about everything you think! I'd love to have a good friendship with you.}\\
\textbf{Login}\\
\textit{Bot: To start talking, I'd like to know your alias. If you have one, write it (without spaces), if not, write: /noAlias}\\
User: /noAlias\\
\textit{Bot: Welcome to the BigHug project! \includegraphics[height=12pt]{figure4.png}. This project aims to build an artificial intelligence that learns to detect different types of disorder and for this, we need people like you to have conversations in a TOTALLY CONFIDENTIAL\includegraphics[height=12pt]{figure5.png}! To participate, you only have to enter an identifying alias with which you want to hide, for example, turtle12 or 110mountain. You can access our website [website] for more information. If you have any questions, we will be happy to help you by emailing [email]. To continue, enter your new alias.}\\
User: croquette13\\
\textbf{General questions}\\
\textit{Bot: Great! Your new alias is: croquette13. Use it to talk to me \includegraphics[height=12pt]{figure12.png} And... hey, how old are you?}\\
User: 14\\
\textit{Bot: How nice! I'm only a few weeks old \includegraphics[height=12pt]{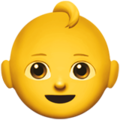} \includegraphics[height=12pt]{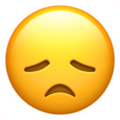}. These days I would love to learn a bit about how you are and how you feel. That will help me to understand \includegraphics[height=12pt]{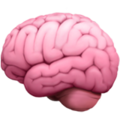} also how I am and how I feel. That's why I'll be asking you some questions. To talk to you... Which gender would you feel more comfortable with (Female or Male)?}\\
User: Female\\
\textbf{Controlled dialogue}\\
\textit{Bot: Choose the topic you would like us to talk about: \\ /Topic0 - Separation \\ /Topic1 - Bullying \\ /Topic2 - Society \\ /Topic3 - Nerves \\ /Topic4 - TOC \\ /Topic5 - State of shock \\ /Topic6 - Fear \\ /Topic7 - Panic \\ /Topic8 - Sadness I \\ /Topic9 - Sadness II \\ /Topic10 - Food I \\ /Topic11 - Food II \\ /Topic12 - Videogames \\ /Topic13 - Places}\\
User: /Topic12\\
\textbf{Triage questions}\\
\textit{Bot: As I don't know much about you, I would like to ask you a few questions before I can chat with you so I can get to know you a little better.}\\
\textit{Bot: Is the time you spend playing video games causing problems with your friends, your family or your studies? (Yes or No)}\\
User: No\\
\textit{Bot: Have you spent a lot of money to improve your skills or appearance in a video game? (Yes or No)}\\
User: No \\
\textit{Bot: From 0 (not at all) to 8 (a lot), how much do you think the time you spend playing games is affecting your life?}\\
User: 1 \\
\textbf{Controlled dialogue taking into account user sensitivity}\\
\textit{Bot: Today I would like to talk about my favourite topic online video gaming! However, in the last year, I've had some problems because of spending so much time playing... I've dropped my grades, I argue more with my parents since I've been playing... I've even stopped meeting friends to spend time playing... I've even stopped meeting up with friends to spend time playing...I love spending time playing video games, how much time do you spend playing video games?}\\
User: None\\
User: I don't play a lot of video games\\
\textbf{Open dialogue}\\
\textit{Bot: Why don't you play video games?}\\
User: I do not have time \\
\textit{Bot: How much time do you have?}\\
User:  Two hours\\
\textit{Bot: Why don't you spend that time playing video games? }\\
User: Because I feel like doing other things\\
\textit{Bot: That is fantastic. What are some of the things you like to do? }\\
User: I like to play sports, read, watch films...\\
\textbf{Semi-controlled dialogue}\\
\textit{Bot: You seem to have many different interests. Do you feel you are able to balance them all, or do you feel that one of them is taking up too much of your time? (if you want to change the subject tell me /changeTopic, if you want me to tell you something else tell me /tellMeAnotherThing, otherwise continue the conversation by chatting with me).}\\
User: /changeTopic \\
\textbf{Controlled dialogue}\\
\textit{Bot: Choose the topic you would like us to talk about: \\ /Topic0 - Separation \\ /Topic1 - Bullying \\ /Topic2 - Society \\ /Topic3 - Nerves \\ /Topic4 - TOC \\ /Topic5 - State of shock \\ /Topic6 - Fear \\ /Topic7 - Panic \\ /Topic8 - Sadness I \\ /Topic9 - Sadness II \\ /Topic10 - Food I \\ /Topic11 - Food II \\ /Topic12 - Videogames \\ /Topic13 - Places}\\
User: /Topic8\\
\textbf{Triage questions}\\
\textit{Bot:  As I don't know much about you, I'd like to ask you a few questions before I can chat with you so I can get to know you a bit better}\\
\textit{Bot: Do you notice that you have been sad or grumpy more days than you were well in the last year?}\\
User: ... \\

\section{Evaluation questionnaire}
\label{sec:evaluation}
The first part of the questionnaire.

\begin{itemize}
    \item Q1. Gender: (a) Male, (b) Female, (c) Other, (d) Prefer not to say.
    \item Q2. Were you interviewed by our psychologists?: (a) Yes, (b) No.
    \item Q3. Why did you decide to join our project?: \textit{(Free response)}
\end{itemize}

The second part of the questionnaire.

\begin{itemize}
    \item Q4. Did you like the fact that the BOT was on the Telegram social network?: (a) Yes, (b) No, (c) I prefer another social network like WhatsApp, (d) I do not care.
    \item Q5. How often have you used the BOT?: \textit{Range 0 (Nothing) to 10 (All days).}
    \item Q6. If you have made little use of the Bot, why is this?: (a) I have not had time, (b) I was not interested in it, (c) I did not like it, (d) I no longer wanted to be part of the project, (e) It has given me a lot of problems, (f) \textit{(Free response)}
    \item Q7. Has the chatbot lived up to your expectations? What did you expect?: \textit{(Free response)}
    \item Q8. If you have had problems with the BOT, what problems have you had?: (a) It gave nonsense answers, (b) It repeated messages, (c) It was very heavy or insistent, (d) It did not understand what I was writing, (e) It did not follow the conversation, (f) I have not had any problems, (g) \textit{(Free response)}.
    \item Q9. What things would you improve about the chatbot?: \textit{(Free response)}
    \item Q10. What did you like most about the chatbot?: \textit{(Free response)}
    \item Q11. What did you like least about the chatbot?: \textit{(Free response)}
    \item Q12. Did you like visualising a ranking of the chatbot's best friends on the web? Did it motivate you to talk to him?: \textit{(Free response)}
    \item Q13. What would have motivated you to talk more with the chatbot?: \textit{(Free response)}
    \item Q14. Tell us your general impressions (optional): \textit{(Free response)}
\end{itemize}

The last part of the questionnaire.

\begin{itemize}
    \item Q15. Is the objective of our project clear to you?: \textit{Range 0 (Nothing) to 5 (A lot).}
    \item Q16. Should we have remembered to use the chatbot more often?: (a) Yes, (b) Maybe, (c) No, (d) \textit{(Free response)}.
    \item Q17. Comments (optional): \textit{(Free response)}
\end{itemize}

\section{Figures}
\label{sec:figures}
\begin{figure}[H]
     \centering
     \begin{subfigure}[b]{0.32\textwidth}
         \centering
         \includegraphics[width=\textwidth]{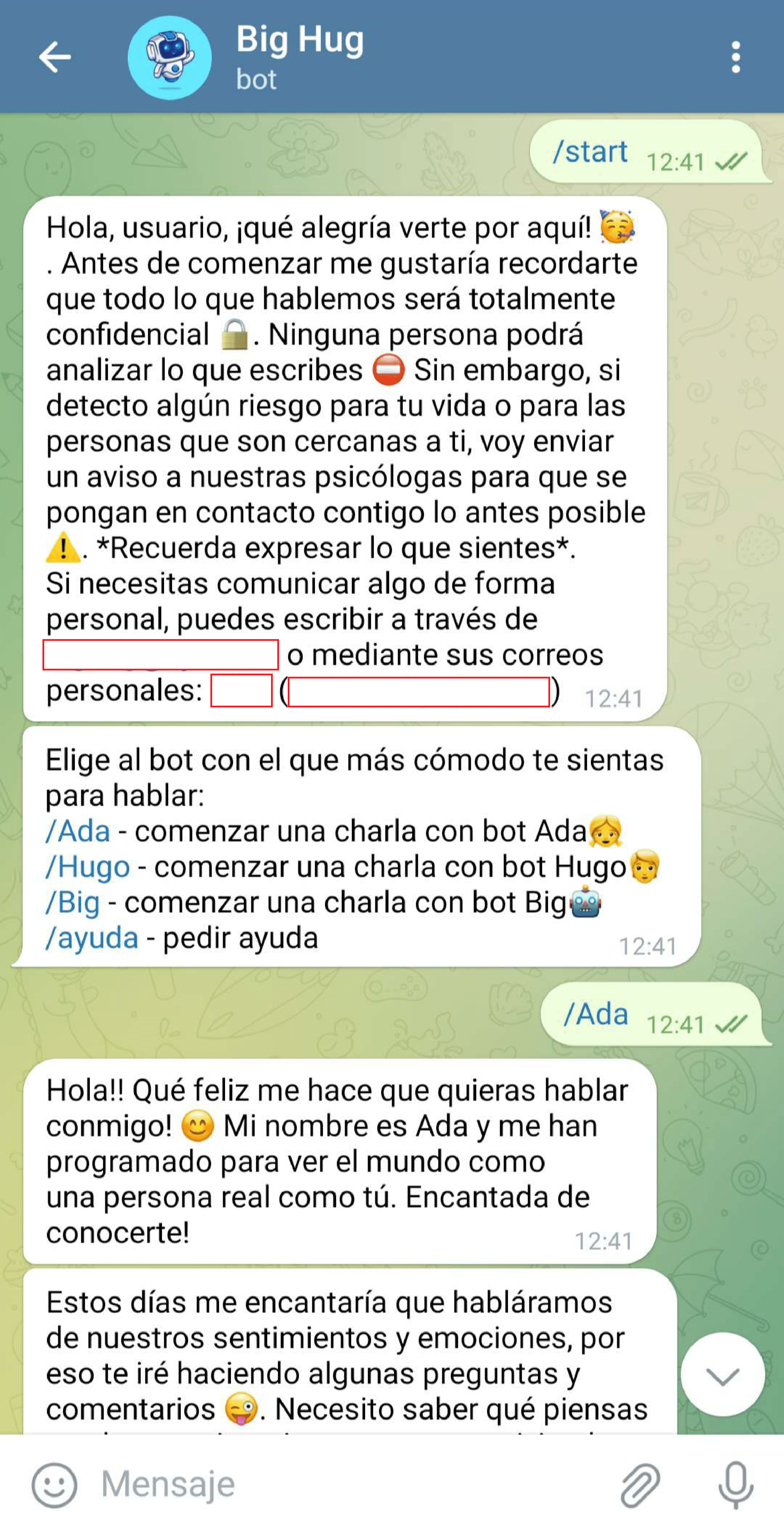}
         \caption{Start of the conversation}
         \label{fig1}
     \end{subfigure}
     \hfill
     \begin{subfigure}[b]{0.32\textwidth}
         \centering
         \includegraphics[width=\textwidth]{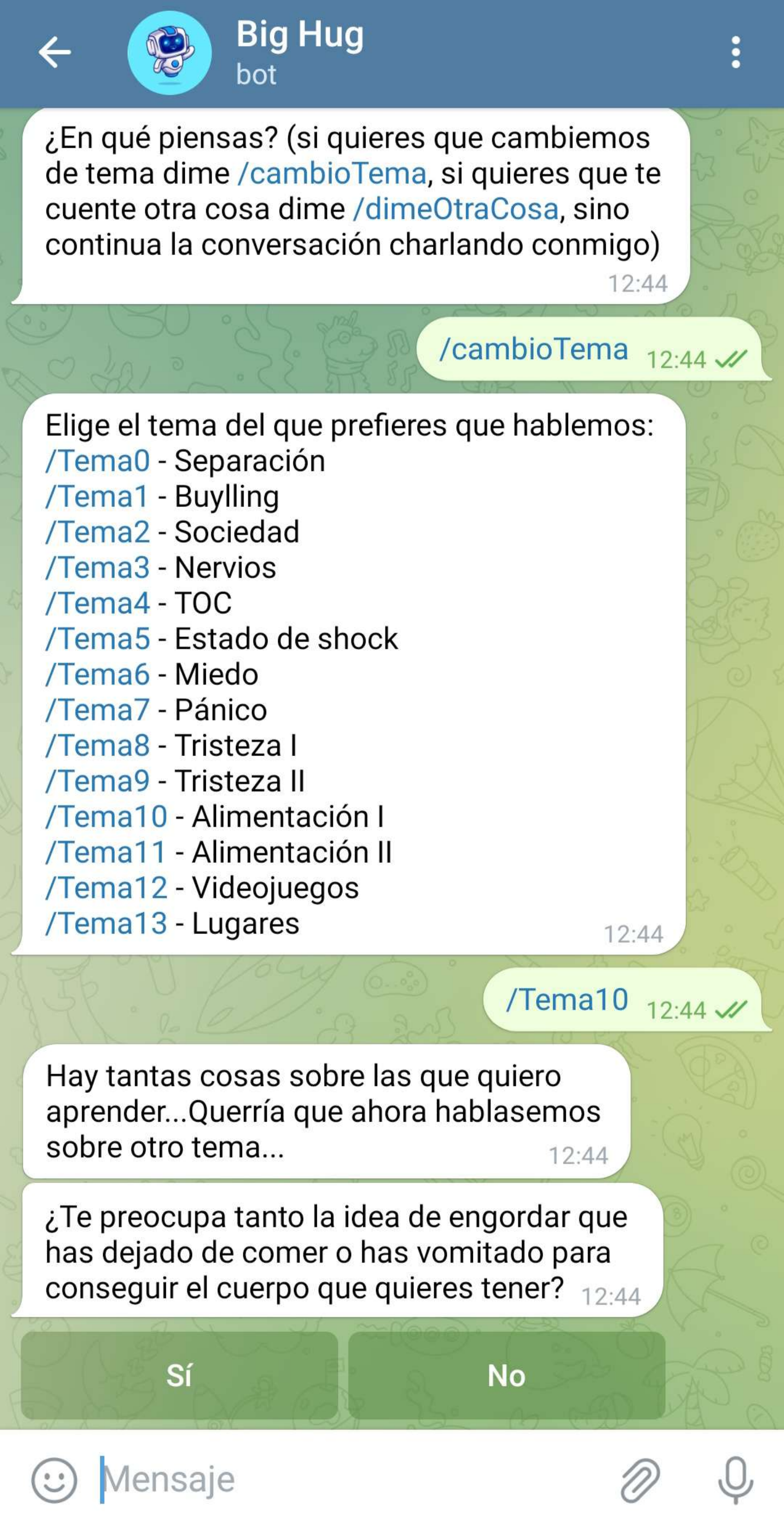}
         \caption{End of the fifth interaction, change of topic and start of triage questions.}
         \label{fig2}
     \end{subfigure}
    \hfill
     \begin{subfigure}[b]{0.32\textwidth}
         \centering
         \includegraphics[width=\textwidth]{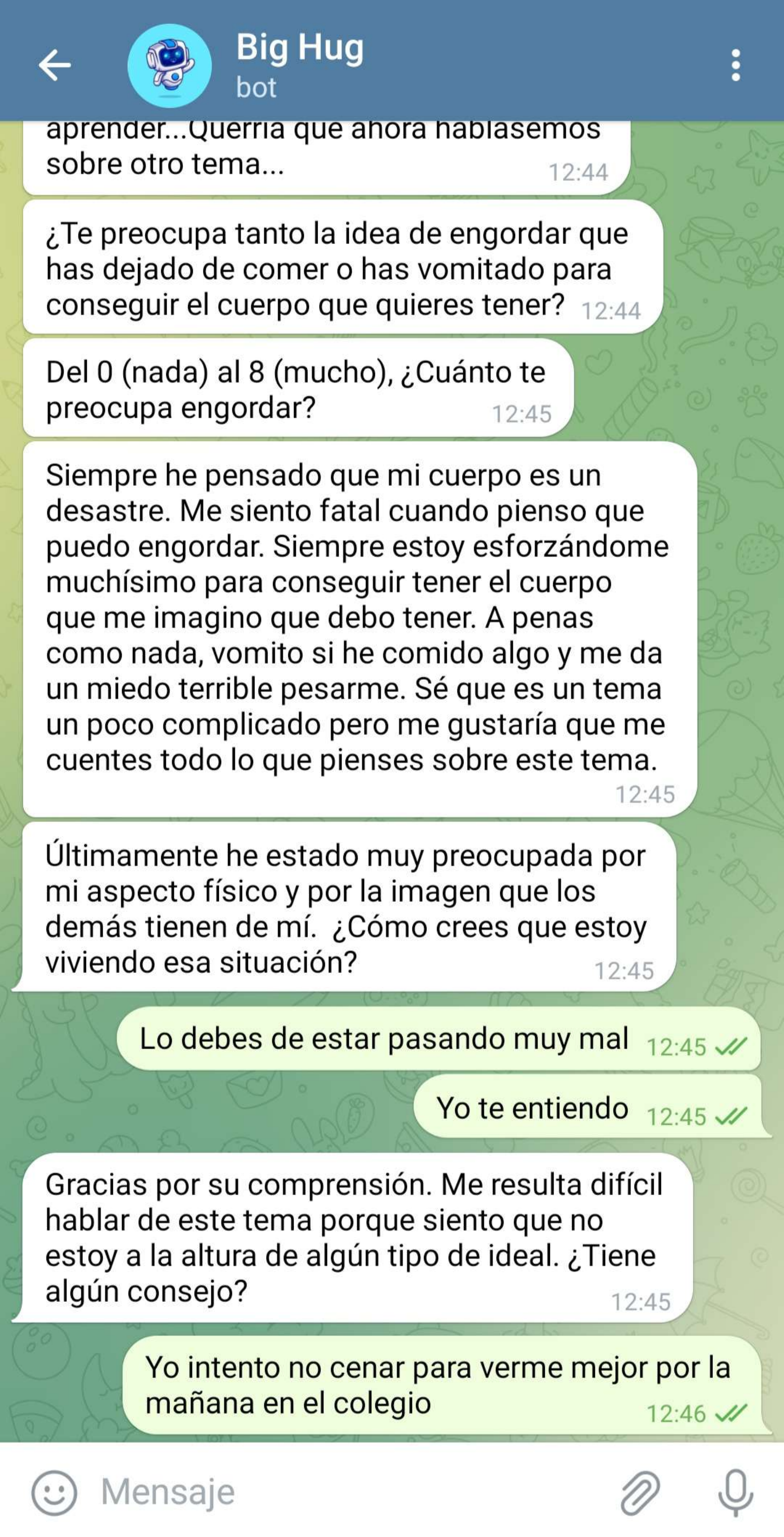}
         \caption{Sample conversation}
         \label{fig3}
     \end{subfigure}
        \caption{Demo of chat interface in Telegram.}
        \label{fig:three graphs}
\end{figure}

Below are some screenshots (Figure \ref{fig:three graphs}) of parts of the conversation with the chatbot on Telegram. 

\bibliographystyle{elsarticlenum}
\bibliography{custom}

\end{document}